\documentclass[a4paper,11pt]{article}
\pdfoutput=1 % if your are submitting a pdflatex (i.e. if you have
\usepackage{jheppub} % for details on the use of the package, please
\usepackage[T1]{fontenc} % if needed 
\usepackage{amsmath}
\usepackage{graphicx}
\usepackage{amssymb}
\usepackage{subfigure}
\usepackage{cancel}
\usepackage{relsize}
\usepackage{bbold}
\usepackage{lineno}
\usepackage{braket}
\usepackage{slashed}
\usepackage{multirow}
\usepackage{placeins}
\usepackage{physics}
\usepackage[utf8]{inputenc}

\definecolor{nicered}{rgb}{0.7,0.1,0.1}
\definecolor{nicegreen}{rgb}{0.1,0.5,0.1}
\definecolor{red}{rgb}{1.0, 0, 0}
\definecolor{niceblue}{rgb}{0,0,0.8}
\hypersetup{colorlinks,citecolor= niceblue,linkcolor= niceblue,urlcolor=niceblue}

\newcommand{\beq} {\begin{equation}}
\newcommand{\eeq} {\end{equation}}
\newcommand{\bea} {\begin{eqnarray}}
\newcommand{\eea} {\end{eqnarray}}
\newcommand{\ba} {\begin{eqnarray*}}
\newcommand{\ea} {\end{eqnarray*}}
\newcommand{\GeV} {\,\text{GeV}}

\newcommand{\hc} {\,\text{h.c.}}
\newcommand{\eps} {\epsilon}
\newcommand{\no}{\nonumber}

\renewcommand{\L} {\mathcal{L}}

\renewcommand{\[}{\left[}

\renewcommand{\(}{\left(}
\renewcommand{\)}{\right)}

\preprint{UCI-TR-2020-13}

\title{Minimal Froggatt-Nielsen Textures}

\author{Marco Fedele$^{a}$,} 
\author{Alessio Mastroddi$^{b,c}$,}
\author{and Mauro Valli$^{d}$}

\affiliation{$^a$Dept.~de F\'{\i}sica Qu\`antica i Astrof\'{\i}sica, Institut de Ci\`encies del Cosmos (ICCUB), 
Universitat de Barcelona, Mart\'i i Franqu\`es 1, E-08028 Barcelona, Spain}
\affiliation{$^b$Dipartimento di Matematica e Fisica, Universit\`a di Roma Tre, I-00146, Rome, Italy} 
\affiliation{$^c$INFN, Sezione di Roma Tre, Via della Vasca Navale 84, I-00146 Rome, Italy}
\affiliation{$^d$Department of Physics and Astronomy, University of California, Irvine, CA 92697-4575 USA}

\emailAdd{marco.fedele@icc.ub.edu}
\emailAdd{alessio.mastroddi@uniroma3.it}
\emailAdd{mvalli@uci.edu}

\abstract{
The flavour problem of the Standard Model can be addressed through the Froggatt-Nielsen (FN) mechanism.
In this work, we develop an approach to the study of FN textures building a direct link between FN-charge assignments and the measured masses and mixing angles via unitary transformations in flavour space. We specifically focus on the quark sector to identify the most economic FN models able to provide a dynamical and natural understanding of the flavour puzzle. 
Remarkably, we find viable FN textures, involving charges under the horizontal symmetry that do not exceed one in absolute value (in units of the flavon charge). 
Within our approach, we also explore the degree of tuning of FN models in solving the flavour problem via a measure analogous to the Barbieri-Giudice one. We find that most of the solutions do not involve peculiar cancellations in flavour space.
}
  
\begin{document} 

\flushbottom
\allowdisplaybreaks

\maketitle

\section{Introduction}
\label{sec:intro}
In recent years much of the focus of the community has been put into understanding the naturalness of the electroweak (EW) scale, namely how to radiatively stabilize the dynamics underlying EW symmetry breaking. 
In absence of the related new-physics signals at the LHC~\cite{Feng:2013pwa,Craig:2013cxa,Panico:2015jxa}, a critical rethinking of the EW hierarchy problem and of the implications in the search for Physics Beyond the SM (BSM) may be the necessary step forward~\cite{Farina:2013mla,
deGouvea:2014xba, Dine:2015xga, Graham:2015cka, Giudice:2017pzm}.

Nevertheless, the prediction of the peculiar flavour structure of the SM Yukawa sector might actually underlie another interesting theoretical problem in virtue of the hierarchy of masses among the three fermion generations, and the different mixing patterns in the quark and lepton sectors~\cite{Weinberg:1977hb,Fritzsch:1999ee}.
Indeed, the quest on the origin and the size of the breaking of the flavour group $\mathcal{G}_{F} = U(3)_{Q} \times U(3)_{u} \times U(3)_{d} \times U(3)_{L} \times U(3)_{e} \times U(1)_{H}$ -- the maximal global symmetry commuting with space-time symmetries, that leaves invariant the gauge-kinetic sector of SM quark ($Q_{i}$,$u_{i}$,$d_{i}$), lepton ($L_i$,$e_i$) and Higgs ($H$) fields ($i=1,2,3$ family index) -- characterizes the so-called \textit{flavour problem} of the SM~\cite{Raby:1995uv,GGRossFlavour}.

In light of the experimental information at disposal on the mass spectrum of SM fermions and the textures shaping the Cabibbo-Kobayashi-Maskawa (CKM)~\cite{Cabibbo:1963yz,Kobayashi:1973fv} and Pontecorvo-Maki-Nakagawa-Sakata (PMNS)~\cite{Pontecorvo:1957qd,Maki:1962mu} matrices, a dynamical explanation of the origin of flavour may provide one of the most convincing calls for BSM physics, see, e.g., the broad reviews in~\cite{Babu:2009fd,Feruglio:2015jfa}, and also the more recent one in~\cite{Feruglio:2019ktm}.

In literature, popular approaches aimed at UV-completing the SM with main focus on the flavour puzzle have extensively relied on the existence of ``flavons'': heavy scalar fields whose vacuum expectation value (VEV) yields spontaneous symmetry breaking of a large enough (discrete or continuous) symmetry group, subgroup of $\mathcal{G}_F$, responsible of the low-energy fermion mass spectrum and the mixing-angle patterns. Flavons may be considered ubiquitous in the context of Grand Unified Theories~\cite{Barbieri:1994kw,Barbieri:1996ww,King:2003rf,King:2005bj,Linster:2018avp,Arias-Aragon:2020bzy} and of stringy UV completions~\cite{Ibanez:1994ig,Kobayashi:2006wq,BerasaluceGonzalez:2012vb,King:2013eh,Chen:2014tpa}. Then, the flavour puzzle may offer a phenomenological handle on those UV theories involving very-high energy dynamics difficult to probe at colliders, that may feature the predicted flavour pattern as a key IR signature of the theory~\cite{Binetruy:1994ru,Binetruy:1996xk}. 

Note that the aforementioned flavon models do not set the only possible framework where the SM flavour puzzle can be solved. Radiative corrections may give a clue on why some Yukawa couplings may be tiny but not identically vanishing, as originally pointed out in refs.~\cite{PhysRevLett.29.388,PhysRevD.7.2457,Barr:1976bk,Balakrishna:1988ks} and modernly revisited, for instance, in~\cite{Crivellin:2011sj,Altmannshofer:2014qha}.
Also, the mixing of SM fermions with heavy resonances arising from strongly interacting sectors, as for the case of theories of partial compositeness, may offer a pragmatic solution to the flavour puzzle~\cite{Kaplan:1991dc,Buras:2011ph,Panico:2016ull}, avoiding the need of horizontal-symmetry breaking.
Most notably, hierarchies without symmetries may follow from embedding the SM in a theory of extra-spatial dimensions. In a 5D extension of the SM, the observed quark and lepton masses and mixings could be the reflection of the geometry of the extra dimension and of the specifics on the localization of fermion and Higgs profiles on the IR brane, see~\cite{ArkaniHamed:1999dc,Gherghetta:2000qt,Huber:2000ie,Kaplan:2001ga} and the more recent study in~\cite{Ahmed:2019zxm}.

In this work, we restrict our considerations to four-dimensional theories and aim at critically reviewing the simplest of the flavon models at hand, in a bottom-up perspective close in spirit to the original work of Froggatt and Nielsen (FN) in~\cite{Froggatt:1978nt}, further expanded in~\cite{Leurer:1992wg,Leurer:1993gy}.  
According to the \textit{FN mechanism}, the SM flavour puzzle is addressed by the introduction of an Abelian flavour symmetry $U(1)_{X}$, spontaneously broken by the VEV $v_{\phi}$ of a single flavon field $\phi$; the flavour structure observed at low energy arises once heavy new degrees of freedom -- the FN messengers~\cite{Calibbi:2012yj} -- properly charged under the horizontal symmetry broken by the flavon, have been integrated out at the high-energy scale $\Lambda > v_{\phi}$.
Such a simple setup has recently gained particular attention in the context of a possible flavour window on the QCD axion and on axion-like particles~\cite{Ema:2016ops,Calibbi:2016hwq,Alanne:2018fns,Bonnefoy:2019lsn,MartinCamalich:2020dfe}; for collider programs dedicated to the flavour problem~\cite{Bauer:2016rxs}; for the vacuum stability of the Higgs potential~\cite{Giese:2019khs}; in connection to the present tensions in $B$ physics~\cite{Falkowski:2015zwa,Barbieri:2019zdz,Bordone:2019uzc}; in relation to the clockwork mechanism~\cite{Kaplan:2015fuy,Giudice:2016yja} for flavour via an inverted FN construction~\cite{Alonso:2018bcg,Sannino:2019sch,Smolkovic:2019jow}; as an unorthodox bridge to the fundamental questions in the physics of the Early Universe~\cite{Berkooz:2004kx,Calibbi:2015sfa,Baldes:2016rqn,Baldes:2016gaf,Lillard:2018zts,Chen:2019wnk,Elahi:2020pxl}.

In our study we will characterize a FN model by the set of charges assigned to the matter fields under the horizontal symmetry $U(1)_{X}$, and by a single perturbative parameter $\epsilon$. We will work within an effective field theory (EFT) approach, and as such we will leave unspecified the details of the UV dynamics that will cure any gauge anomaly naively present in the $X$ charge assignments considered at low energy~\cite{PRESKILL1991323}. 
Our focus here is explicitly devoted to the quark sector, where we try to assess in detail the minimal amount of theoretical assumptions needed to precisely reproduce the quark-mass hierarchies and the high hierarchical structure of the CKM matrix in a natural manner.
The main points and novelties of the present paper are the followings:
\begin{itemize}
    \item We offer a new method that guarantees the exact reproduction of observed masses and mixing and a close inspection of the size of the dimensionless couplings involved in FN models, supposedly $\mathcal{O}(1)$. The novelty is to start from a basis where inputs are indeed the precise measurements of fermion masses and mixing angles, an then exploit unitary rotations in $\mathcal{G}_{F}$ to map SM fields into eigenstates of the new FN interactions;
    \item We introduce a fine-tuning measure to keep track of peculiar cancellations in flavour space that would weaken the goodness of a FN model resolving the flavour puzzle;
    \item We provide a bottom-up exploration of the possible minimal charge assignments in FN models via an EFT approach, reproducing quark textures and investigating the role of fine tuning in flavour space.
    %\item We comment on a new potential hierarchy problem arising in the class of minimal FN constructions, concluding de facto that a natural flavon in these models should be linked to heavy new dynamics not so far from the EW scale.
\end{itemize}

\noindent The present paper is organized as follows: In section~\ref{sec:FN_EFT} we review the FN mechanism in its EFT incarnation; Section~\ref{sec:how_to_fit_data} is devoted to introduce in detail our new approach to the study of FN models and it also includes a discussion on the fine-tuning measure for the SM flavour puzzle; Section~\ref{sec:results} contains the main results of our study; In section~\ref{sec:conclusions} we leave our ending comments on the possible interesting future directions to pursue. 

\section{FN in the EFT formulation}
\label{sec:FN_EFT}
Let us briefly review in this section the FN mechanism using the language of EFT, and set the relevant notation adopted in the rest of the paper.
The SM Lagrangian of the quark Yukawa sector can be written as follows:
\beq\label{eq:Y_SM}
-\L_{\rm SM}^{Y^{u,d}} = Y^u_{ij} \, \bar{Q}_i \widetilde{H} u_j + Y^d_{ij} \, \bar{Q}_i H d_j + \hc \ ,
\eeq
where $Q$ is the left-handed quark $SU(2)_L$ doublet, $d$ and $u$ are the right-handed down-type and up-type quark singlets, respectively;  
$H$ is the SM Higgs $SU(2)_L$ doublet and $\widetilde H \equiv i\sigma_2 H^*$; $i,j = 1,2,3$ are generation indices, and sum over repeated indices is understood. The gauge-kinetic terms of the fermion fields are invariant under the flavour symmetry group $U(3)_{Q} \otimes U(3)_{u} \otimes U(3)_{d} \subset \mathcal{G}_{F}$. In particular, a field transformation involving the $ 3 \times 3 $ unitary matrices $V_{Q,u,d}\,$ so that: 
\begin{eqnarray}
Q_i  & \ \ \to \ \ & (V_{Q})_{i j} \, Q_j  \ , \nonumber \\
u_i  & \ \ \to \ \ & (V_{u})_{i j} \, u_j  \ , \\
d_i  & \ \ \to \ \ & (V_{d})_{i j} \, d_j  \ , \nonumber 
\end{eqnarray}
leaves invariant the gauge-kinetic term. Then, the Yukawa sector in Eq.~($\ref{eq:Y_SM}$) provides an explicit breaking of the flavour symmetry group down to the Abelian global symmetry $U(1)_B$, which implies the accidental conservation of baryon number, broken in the SM theory only at the non-perturbative level~\cite{tHooft:1976rip}.

In general, the explicit values of the entries in the Yukawa matrices $Y^u$ and $Y^d$ depend on the basis chosen for the quark fields. With an appropriate choice of the matrices $V_{Q},V_{u},V_{d}$ one can always bring the Lagrangian of Eq.~\eqref{eq:Y_SM} in the following form
 \beq\label{eq:Y_SM_diag}
-\L_{\rm SM}^{Y^{u,d}} = \hat{y}^u_{ij} \, \bar{Q}_i \widetilde{H} u_j + (V_{\rm CKM} \, \hat{y}^d)_{ij}  \, \bar{Q}_i H d_j + \hc \ ,
\eeq
where $V_{\rm CKM}$ corresponds to the CKM matrix and 
\beq
\hat y^d = {\rm diag} ( y_d, y_s, y_b )\,, \qquad\qquad \hat y^u = {\rm diag} ( y_u, y_c, y_t )\,,
\eeq
with $y_q=\sqrt{2}m_q/v_{H}$, $m_{q}$ the mass of the quark $q$, and $v_{H} = 246$~GeV, the VEV of the Higgs. The choice of the basis leading to Eq.~\eqref{eq:Y_SM_diag} may turn out to be particularly convenient if we would like to have a direct handle on the 18 complex entries of the Yukawa matrices in terms of the observed values of quark masses and mixing parameters in the CKM, since $Y^u=\hat{y}^u$ and $Y^d= V_{\rm CKM} \, \hat{y}^d$. In the following we refer to this specific choice -- a useful starting point for the present analysis -- as the \textit{up-aligned basis}.\footnote{An equivalent convenient choice could be the \textit{down-aligned basis}, where $Y^u=V^{\dagger}_{\rm CKM}\hat{y}^u$ and $Y^d=  \hat{y}^d$.}

Let us now discuss the FN mechanism. 
We can reconsider Eq.~\eqref{eq:Y_SM} from the point of view of an EFT invariant under a global $U(1)_X$ transformation that distinguishes fermion families. The rephasing of SM fields under the action of $U(1)_X$ corresponds to:\footnote{A slight generalization of this set would also include the Higgs field, here set to $X_{H}=0$ for simplicity.}
\begin{eqnarray}
Q_j & \ \to \ & e^{i \, \theta X_{Q_j}} \, Q_j \ ,\nonumber \\
u_j & \ \to \ & e^{i \, \theta X_{u_j}} \, u_j \ , \\
d_j & \ \to \ & e^{i \, \theta X_{d_j}} \, d_j \ , \nonumber
\end{eqnarray}
where $\theta \in [0,2\pi]$ is the continuous parameter of the global Abelian transformation and $X_{Q_j,u_j,d_j}$ are the $U(1)_{X}$ charges of the various quark fields with generation index $j$. At the renormalizable level, Yukawa terms in the Lagrangian are forbidden as long as $X_{Q_i} \neq X_{u_j}$ or $X_{Q_i} \neq X_{d_j}$ for any pair $i , j$. As anticipated in the Introduction, to have non-zero values for all the elements of the Yukawa matrices, we can introduce a scalar field $\phi$ with $U(1)_{X}$ charge, that we will set to $X_{\phi}=1$ without loss of generality. According to the charge assignments of the fields illustrated above, we can now write down the following terms
\begin{equation}
\label{eq:EFT_Lagrangian}
\mathcal{L}_{\textrm{FN-EFT}} \supset
\begin{cases}
\ c^u_{ij} \, \bar{Q}_i \widetilde{H} u_j \left( {\phi/\Lambda} \right)^{X_{Q_i} - X_{u_j}} +\hc  &\qquad X_{Q_i} - X_{u_j} \geq 0 \ , \\
\ c^d_{ij} \, \bar{Q}_i H d_j \left( {\phi/\Lambda} \right)^{X_{Q_i} - X_{d_j}}+\hc  &\qquad X_{Q\,i} - X_{d_j} \geq 0   \ , \\
\ c^u_{ij} \, \bar{Q}_i \widetilde{H} u_j \left( {\phi^{\dagger}/\Lambda} \right)^{X_{u_j}-X_{Q_i} }+\hc   &\qquad X_{Q_i} - X_{u_j} \leq 0  \ , \\
\ c^d_{ij} \, \bar{Q}_i H d_j \left( {\phi^{\dagger}/\Lambda} \right)^{X_{d_j}-X_{Q_i}}+\hc   &\qquad X_{Q_i} - X_{d_j} \leq 0  \ , \\ 
\end{cases}
\end{equation}
responsible for the Yukawa terms of Eq.~\eqref{eq:Y_SM} once the flavon $\phi$ acquires a VEV along its real component:\footnote{Note that this can be always achieved by a proper redefinition of the complex scalar field $\phi$.} $\langle \phi \rangle = \langle \phi^{\dagger}  \rangle= v_\phi \neq 0$. 
Eq.~\eqref{eq:EFT_Lagrangian} provides the most general formulation of the EFT Lagrangian invariant under the SM gauge symmetry and the global group $U(1)_X$ allowing for
non-renormalizable operators suppressed by a cut-off scale $\Lambda$ and involving a single flavon field $\phi$.

Therefore, the induced EFT Lagrangian stemming from the FN mechanism reads as
\beq
\label{eq:Y_FN}
-\L_{\textrm{FN-EFT}} = c^u_{ij} \, \eps^{n^u_{ij}} \, \bar{Q}_i \widetilde{H} u_j  + c^d_{ij} \, \eps^{n^d_{ij}} \, \bar{Q}_i H d_j + \hc \ ,
\eeq
where the following definitions have been introduced:
\begin{eqnarray}
\label{eq:FN_basis}
\eps & \ \equiv \ & v_\phi/\Lambda \ , \\
\label{eq:FN_nu}
 n^u_{ij} & \ \equiv \ & | X_{Q_i} - X_{u_j} | \ ,  \\
\label{eq:FN_nd}
 n^d_{ij} & \ \equiv \ & | X_{Q_i} - X_{d_j} |  \ . 
\end{eqnarray}
Inspecting Eq.~\eqref{eq:Y_FN}, we can easily see how the mechanism to explain the SM flavour structure can be at work: if $\epsilon < 1$, hierarchical structures in the Yukawa couplings can be generated through the different powers of $\eps^{n^u_{ij}}$ and $\eps^{n^d_{ij}}$ while the coefficients $c^u_{ij} $ and $c^d_{ij}$ can be naturally of the same order, i.e. not hierarchical and of $\mathcal{O}(1)$ in size. In the following we refer to the \textit{FN basis} as the one where such a mechanism is manifest.\footnote{In other words, the FN basis is the one where all the fields $\psi_i$ have a well-defined transformation property under the action of $U(1)_X$, $\psi_i \to e^{\theta X_{\psi_i}} \psi_i$, i.e. eigenstate of the new FN interactions.}

Before concluding this section, a few remarks are in order:
\begin{itemize}
    \item Eq.~\eqref{eq:Y_SM_diag}. and Eq.~\eqref{eq:Y_FN} describe the quark Yukawa Lagrangian in two different basis: this simple observation is the culprit of the discussion carried out in the next section;
    \item Eq.~\eqref{eq:Y_FN} shows that from the low-energy point of view, it is sufficient to know the expansion parameter $\epsilon$ and the matrices $n^{u,d}_{ij}$ (up to $\mathcal{O}(1)$ coefficients) to describe the Yukawa sector in the FN picture; however, for a fixed value of the expansion parameter, different sets of charge assignments $X_{Q_i,u_i,d_i}$ yielding the same $n^{u,d}_{ij}$ entries, characterize distinct FN models in the UV;
    \item Charging also the Higgs field under $U(1)_X$ would simply correspond in a plus (minus) shift in all the entries of $n^{d}_{ij}$ ($n^{u}_{ij}$) of amount $X_{H}$; hence, the generated hierarchy in masses and mixing is primarily controlled by the fermion charge assignment;
    \item Eq.~\eqref{eq:EFT_Lagrangian} does not assume couplings to $\phi$ or $\phi^{\dagger}$ only: such a UV restriction would be a typical outcome of supersymmetric extensions of the SM  due to the holomorphic property of the superpotential~\cite{Leurer:1992wg,Leurer:1993gy}; a supersymmetric version of Eq.~\eqref{eq:Y_FN} would also need a second Higgs doublet at work, implying a dedicated inspection of the role of the misalignement between the VEV of the two Higgs fields in the analysis of the flavour puzzle~\cite{Dudas:1995yu,Dudas:1996fe,Dery:2016fyj}. In our study we do not consider this class of models, that would offer a  generalization of Eq.~\eqref{eq:Y_FN}.
\end{itemize}

\section{How FN theories confront data}
\label{sec:how_to_fit_data}
Let us now move our discussion to the most relevant point of the paper, namely how a generic FN model yielding at low energy the structure highlighted in Eq.~\eqref{eq:FN_basis}, should be discriminated by the dataset of interest, namely the six quark masses, together with the three mixing angles and the CP-violating phase of the CKM matrix.

\subsection{A customary approach}
In the inspection of the quark sector, in literature it has been often the case of relying on approximate relations that put under the spotlight the strong hierarchies among quark-mass ratios and the CKM entries. This fact induced some authors to pinpoint specific hierarchical structures for the Yukawa matrices $Y^{u,d}_{ij}$, see e.g. ref.~\cite{Hall:1993ni}. Along these lines, a very popular choice is to identify the Cabibbo angle $ \lambda = \sin \theta_{c} \simeq 0.22$ with the expansion parameter $\epsilon$, and then obtain a description of quark masses and mixings according to:
\begin{eqnarray}
\label{eq:approx_pattern}
 y_d \ & \sim & \ \lambda^6 \ ,  \
 y_s \  \sim  \ \lambda^4 \ , \
 y_b \ \sim \ \lambda^2 \ ,  \
 y_u \ \sim \ \lambda^7 \ , \
 y_c \ \sim \ \lambda^3 \ , \
 y_t \ \sim \ \lambda^0 \ ,  \\
|V_{ud}| & \ \sim \ & |V_{cs}| \sim |V_{tb}|  \sim \lambda^0 \ , \
|V_{us}|  \sim |V_{cd}| \sim \lambda \ , \
|V_{cb}| \sim |V_{ts}| \sim \lambda^2 \ , \
|V_{ub}| \sim |V_{td}| \sim \lambda^3 \  \nonumber . 
\end{eqnarray}
The approximate relations reported above represent a valuable benchmark for a top-down approach that aims at qualitatively explaining the SM flavour puzzle. On the other hand, it is quite easy to single out a FN model that from the bottom-up point of view -- up to multiplicative factors of $\mathcal{O}(1)$ -- is able to reproduce the relations highlighted in Eq.~\eqref{eq:approx_pattern}. For instance, assigning $X_{Q_{1,2,3}} = \{3,2,0\}$, $X_{u_{1,2,3}} = \{-4,-1,0\}$, $X_{d_{1,2,3}} = \{-3,-2,-2\}$, together with $\epsilon \sim \lambda$, one would accomplish the goal of reproducing the pattern in Eq.~\eqref{eq:approx_pattern}~\cite{Feruglio:2015jfa,Baldes:2016gaf}.

Needless to say, a more quantitative investigation of the SM flavour problem may be highly desirable. Within the FN mechanism, this requires to go well beyond the order-of-magnitude scalings of Eq.~\eqref{eq:approx_pattern}. In fact, one may attempt to reproduce the value of quark masses (evaluated at the high-energy scale under scrutiny) more precisely, and also succeed in describing the detailed textures of the CKM matrix~\cite{Grossman:2020qrp}. The latter, in fact, does not have entries that in absolute value really yield a symmetric matrix, with the size of the mixing of the first and third generation badly breaking such (widespread) approximation, see the most updated results on unitarity-triangle analyses in~\cite{Charles:2015gya,Bona:2017gut,UTfit2018}.

In a more ambitious endeavour on the assessment of a solution to the SM flavour puzzle, one may perform a precise fit to the observed masses and mixing angles in the theoretical framework at hand. On mathematical grounds, one could formulate for the purpose an optimization problem with the following cost function:
\begin{equation}
    \label{eq:chi2_standard}
    \chi^2_{O} = \sum_{K} \left( \frac{\langle O_K \rangle - \widehat{O}_K}{\Delta O_K} \right)^2 \ ,
\end{equation}  
where $O_K$ stands for the observable with measured value $\langle O_K \rangle \pm \Delta O_K$, and theory prediction $\widehat{O}_K$, with $ K = 1, \dots, 10 $ running over the six quark masses, the three CKM mixing angles and, eventually, also on the CP-violating phase. 
In order to study the SM flavour puzzle in the context of the FN mechanism, one may ideally proceed as follows:
\begin{itemize}
    \item Specify a model, fixing the set of nine fermion charges under the global $U(1)_X$;
    \item Use Eq.~\eqref{eq:Y_FN} to write down $\widehat{O}_K$ as a function of 18 complex parameters, characterizing the entries of $c^{u,d}_{ij}$ matrices, and also of the perturbative parameter $\epsilon$;
    \item Minimize $\chi^2_{O}$ to find the optimal values for $c^{u,d}_{ij}$ and $\epsilon$ that reproduce the masses and mixing pattern of the quark sector of the SM;
    \item Consider to accept or reject the FN model on the basis of the textures found in $c^{u,d}_{ij}$: a successful FN model should feature non-hierarchical $\mathcal{O}(1)$ entries for $|c^{u,d}_{ij}|$.
\end{itemize}  

Note that the problem so formulated involves 37 real parameters to be determined from 10 measurements.
% Note that given the degeneracy of the problem, namely 37 real parameters to be determined from only 10 measurements, further theoretical input would be needed in order to perform a meaningful minimization of $\chi^2_{O}(c^{u,d}_{ij},\epsilon)$ at the practical level. 
% A simple option would be to drastically reduce the number of parameters involved leaving the discussion on CP violation aside, i.e. focusing only on masses and mixing angles.
To drastically reduce the number of parameters involved, one could leave the discussion on CP violation aside, i.e. focusing only on masses and mixing angles.
Furthermore, one may also set $\epsilon$ to a reasonable benchmark of interest (e.g., the Cabibbo angle highlighted in the relations of Eq.~\eqref{eq:approx_pattern}), instead of inferring it from data. Then, one would end up with 18 parameters to be extracted from a fit to 9 measurements, assessing the goodness of the flavour model on the basis of how many of these 18 fitted coefficients would turn out to be $\mathcal{O}(1)$. 
% However, despite the simplification of the initial problem, several flat directions in parameter space would still develop in the procedure of minimizing Eq.~\eqref{eq:chi2_standard}. Hence, some further well-motivated theoretical handle would be advantageous. 

% Along the lines of what presented recently in ref.~\cite{Linster:2018avp}, such an improvement can be provided by introducing an additional weight to $\chi^2_{O}$ that takes into account a notion of distance for the coefficients from the expected $\mathcal{O}(1)$ value; for instance, one may consider:
Along the lines of what presented recently in ref.~\cite{Linster:2018avp}, a further improvement of this method can be provided by introducing an additional weight to $\chi^2_{O}$ that takes into account a notion of distance for the coefficients from the expected $\mathcal{O}(1)$ value; e.g., one may consider:
\begin{eqnarray}
    \label{eq:chi2_stand_improved}
    \chi^2_{\rm tot} & \ \ = \ \ &  \chi^2_O + \chi^2_{\mathcal{O}(1)} \ , \\
    \chi^2_{\mathcal{O}(1)}  & \ = \ & \sum_{q = u,d} \,\sum_{i,j=1}^3 \left(\frac{|c^{q}_{ij}| - \mu_{c}}{\sigma_{c}} \right)^2 \nonumber \ ,
\end{eqnarray}
implying a normal distribution for the absolute value of the coefficients $c^{u,d}_{ij}$ and standard deviation $\sigma_c$. A reasonable choice may be then to set the mean to unity and, pending on the strictness of the ``$\mathcal{O}(1)$ requirement'', $\sigma_c$ could be matching, e.g., 10\% level.

The improved cost function in Eq.~\eqref{eq:chi2_stand_improved} allows to turn the original optimization problem into an overdetermined system, with 27 independent constraints potentially nailing the global minimum in the 18-dimensional parameter space (or 19-dimensional one if the FN scale-ratio $\epsilon$ is not fixed a priori). Note, however, that the goodness of this approach in the end critically depends on the choice of $\sigma_{c}$, that establishes which of the two terms in Eq.~\eqref{eq:chi2_stand_improved} weights the most in the minimization procedure. Moreover, the original problem at hand got the main simplification from the requirement of dealing only with real dimensionless coefficients. 
% Eventually, as we are going to show in what follows, a survey of FN models according to the algorithm above would not exploit the symmetries of the problem, namely the $U(3)^3$ rotations in flavour space, that allows one to pin down the actual degrees of freedom relevant for the study of the SM flavour puzzle.
In the following, we are going to discuss how a different approach would allow one to simplify the problem at hand without relying on any assumption of this sort. 
In particular, within such an approach we are going to show how $U(3)^3$ rotations in flavour space can be generally exploited in order to survey FN models very efficiently.

\subsection{A novel approach}
Let us now consider an alternative approach to Eq.~\eqref{eq:chi2_stand_improved} that allows a general exploration of the flavour problem within the FN EFT where we aim at:
\begin{itemize}
\item Explaining the pattern of SM quark masses and mixing, to be precisely reproduced;
\item Obtaining the size of the dimensionless coefficients $c^u_{ij} $ and $c^d_{ij} $ in a \textit{natural range};
\item Avoiding any fine tuning possibly occurring among the parameters involved.
\end{itemize}

As a starting point, let us introduce the approximation $\Delta O_K/\langle O_{K}\rangle \ll 1$ as our working hypothesis, namely let us focus only on the central values for the available measurements of quark masses and CKM mixing parameters. Given the $\mathcal{O}(1)$ characterization of the coefficients in Eq.~\eqref{eq:Y_FN}, such an approximation should be considered reasonable in the present context: charm, bottom and top quarks currently show an uncertainty on the determination of the mass of the percent level, and such a precise determination holds true for CKM angles as well, while $\Delta O_K/\langle O_{K}\rangle$ floats around $\mathcal{O}(10\%)$ for the lightest three quarks and the CKM phase~\cite{PhysRevD.98.030001}.

Given the set of measurements $\langle O_{K}\rangle$, $K=1,\dots,10\,$, we can easily construct the quark Yukawa sector in the up-quark aligned basis reported in Eq.~\eqref{eq:Y_SM_diag}. Then, we observe that the aligned basis and the FN basis are related by a transformation of $U(3)^3$. In other words, we know that there exist three unitary matrices $V_Q$, $V_u$ and $V_d$ such that: 
\beq
\label{eq:FN_rotation}
\(V_Q^{\dagger} \, \hat{y}^u \, V_u\)_{ij}= c^u_{ij} \, \eps^{n^u_{ij}} \ \ \ \ , \ \ \ \(V_Q^\dagger \,  V_{\rm CKM} \hat{y}^d  \, V_d\)_{ij} = c^d_{ij} \, \eps^{n^d_{ij}} \ \ .
\eeq
Consequently, for a set of assignments of FN charges and for a given value of $\epsilon$, we can always rewrite the dimensionless coefficients $c^u_{ij}$ and $c^d_{ij}$ in terms of the above unitary matrices.

The statement above is the key observation of the present study.
Indeed, using Eq.~\eqref{eq:FN_rotation} the three unitary matrices can completely specify the values of the coefficients $c^u_{ij}$ and $c^d_{ij}$. For every point in the $U(3)^3$ parameter space we can then compute the values of the elements of $c^u_{ij}$ and $c^d_{ij}$. In this way, the problem of addressing the SM flavour puzzle via the FN mechanism can be formulated in terms of the parameters that span the $U(3)^3$ flavour space of the quark sector. In particular, from the compact ranges of the $3 \times 9 = 27$ independent parameters one can generate all the possible existing $V_{Q,u,d}$ matrices. 
Note that in the customary approach previously presented, for a given set of coefficients $c^{u,d}_{ij}$, these rotation matrices are uniquely determined by the diagonalization procedure involved in the computation of $\widehat{O}_{K}$, contrary to what proposed in the new formulation of the problem.

Let us now make a few observations directed at further simplifying the analysis without any loss of generality. In first place, given $y_{t} \simeq 1$,\footnote{The approximation holds also for the running of the top-quark mass from the EW scale up to the TeV.} we can establish a relation between the FN charges in (what-would-be) the top-quark sector:
\beq
\label{eq:FN_charge_top}
n^u_{33} \simeq 0 \ \Rightarrow \  {X_Q}_3 = {X_u}_3\,.
\eeq
Moreover, the accidental baryon symmetry $U(1)_{B}$ allows us to remove one of the physical assignments for the FN charges. Hence, under the approximation highlighted in Eq.~\eqref{eq:FN_charge_top} we can safely set:
\beq
\label{eq:zero_charges}
{X_Q}_3 = {X_u}_3 = 0\,.
\eeq
As a result, only seven $U(1)_X$ charges out of the initial nine ones are actually independent and need to be assigned to characterize the construction of the FN EFT in Eq.~\eqref{eq:Y_FN}.

In order to directly compare Eq.~\eqref{eq:Y_SM} with Eq.~\eqref{eq:Y_FN}, we need to introduce three distinct $U(3)$ transformations as already illustrated in Eq.~\eqref{eq:FN_rotation}. An element $V \in U(3)$ can be uniquely defined by a transformation involving three angles, $\theta_{1,2,3}$, and six phases, $\delta_{1,\dots,6}$; in the fundamental representation this one can be constructed as follows~\cite{Rasin:1997pn,Fritzsch:1997st}:
\beq
\label{eq:Euler_U3}
 V =
\left(\begin{array}{ccc}
1&0&0 \\
0 & \ e^{i\delta_2} & 0\\
0 & 0 & e^{i\delta_3}
\end{array}\right)
\left(\begin{array}{ccc}
c_1c_2 & c_1s_2 & s_1e^{-i\delta_1}\\ 
-c_{3}s_{2} -c_{2}s_{1}s_{3}e^{i\delta_1}&c_{2}c_{3}-s_{1}s_{2}s_{3}e^{i\delta_1}& c_{1}s_{3}\\ 
s_{2}s_{3} -c_{2}c_{3}s_{1}e^{i\delta_1}&-c_{2}s_{3}-c_{3}s_{1}s_{2}e^{i\delta_1}&c_{1}c_{3}
\end{array}\right)
\left(\begin{array}{ccc}
e^{i\delta_4}&0 &0\\
0 & e^{i\delta_5}&0\\
0&0&e^{i\delta_6}
\end{array}\right)\,,
\eeq
where the shorthands $c_{1,2,3} \equiv \cos \theta_{1,2,3}$, $s_{1,2,3} \equiv \sin \theta_{1,2,3}$ have been adopted. 
Hence, a priori, the three unitary transformations in Eq.~\eqref{eq:FN_rotation}, $V_{Q,u,d}$, would indeed involve a total of 27 independent parameters. However, note that each of the nine quark fields can be redefined under a $U(1)$ transformation, innocuous on the gauge-kinetic sector of the SM Lagrangian, with $ U(1)^9 \subset \mathcal{G}_{F}$. This allows one to eliminate three of the six phases present in each of $V_{Q,u,d}$, which consequently can take the reduced form:
\beq
V = 
\left(\begin{array}{ccc}
\label{eq:Euler_U3_3phases}
c_{1}c_{2} \ & \ c_{1}s_{2} & \ s_{1}e^{-i\delta_1}\\ 
-c_{3}s_{2}e^{i\delta_2} -c_{2}s_{1}s_{3}e^{i(\delta_1+\delta_2)}& \ c_{2}c_{3}e^{i\delta_2}-s_{1}s_{2}s_{3}e^{i(\delta_1+\delta_2)}& \ c_{1}s_{3}e^{i\delta_2} \\
s_{2}s_{3 }e^{i\delta_3} -c_{2}c_{3}s_{1}e^{i(\delta_1+\delta_3)}& \ -c_{2}s_{3}e^{i\delta_3}-c_{3}s_{1}s_{2}e^{i(\delta_1+\delta_3)}& \ c_{1}c_{3}e^{i\delta_3}
\end{array}\right)\,.
\eeq

Therefore, the number of degrees of freedom characterizing the problem corresponds only to $3 \times 3 = 9$ mixing angles, varying in the compact interval [0,$\pi$], and $3 \times 3 = 9$ phases, spanning the range [0,$2\pi$], yielding a total of 18 independent parameters.
 
We can now present a step-by-step analysis of the SM flavour puzzle in FN theories within such a general setting.
As a starting point, one randomly generates a set of angles $\theta_{1,2,3}^{\,Q,u,d}$ and phases $\delta_{1,2,3}^{\,Q,u,d}$, that identify a point in the  $U(3)^3$ quark flavour space. 
Within the FN EFT under consideration, one should proceed characterizing the FN model. As previously mentioned, this can be done fixing a set of FN charges, defining $n^{u,d}_{ij}$ in Eq.~\eqref{eq:FN_basis}, together with $\epsilon$, that could be randomly extracted in the interval $(0,1)$ being an expansion parameter. As a second step, exploiting the precise measurements of $\hat{y}^{u,d}$ and $V_{\rm CKM}$, one can evaluate $c^{u,d}_{ij}$ inverting the relations presented in Eq.~\eqref{eq:FN_rotation}:
\begin{eqnarray}
\label{eq:FN_coeffs}
c^u_{ij}( \epsilon, \theta_{1,2,3}^{\,Q,u,d}, \delta_{1,2,3}^{\,Q,u,d} \,) & \ = \ & \(V_Q^{\dagger} \, \hat{y}^u \, V_u\)_{ij} \, / \,  \eps^{n^u_{ij}} \ \ , \nonumber \\
c^d_{ij}( \epsilon, \theta_{1,2,3}^{\,Q,u,d}, \delta_{1,2,3}^{\,Q,u,d} \, ) & \ = \ & \(V_Q^\dagger \,  V_{\rm CKM} \, \hat{y}^d  \, V_d \)_{ij} \,  / \, \eps^{n^d_{ij}} \ ,
\end{eqnarray}
where we have highlighted that for a given FN charge assignment, $c^{u,d}_{ij}$ are now considered as functions of the perturbative parameter $\eps$ and of the \textit{18 nuisance parameters} corresponding to 9 independent angles and phases. 
This formulation may be advantageous from the following point of view: while inferring the value of $\epsilon$ from data retains a clear phenomenological relevance, angles and phases can be easily spanned in compact intervals to extensively chart the parameter space and assess whether $c^{u,d}_{ij}$  entries may turn out to be $\mathcal{O}(1)$ coefficients.

So, as a last step, one can introduce the new cost function:
\begin{equation}
    \label{eq:FN_cost_function}
    \chi^2_{\rm FN} = \sum_{i,j=1}^{3} \left( |c^u_{ij}( \epsilon, \theta_{1,2,3}^{\,Q,u,d}, \delta_{1,2,3}^{\,Q,u,d} \,)| -1 \right)^2 + \left( |c^d_{ij}( \epsilon, \theta_{1,2,3}^{\,Q,u,d}, \delta_{1,2,3}^{\,Q,u,d} \,)| -1 \right)^2 \ ,
\end{equation}
and minimize it with respect to $U(3)^3$ angles, phases, and the expansion parameter $\epsilon$.

Finally, one should establish a criterion of acceptance for a natural range of $c^{u,d}_{ij}$, that will characterize them as  $\mathcal{O}(1)$ parameters. For instance, given some departure $\Delta x$ from unity, one might retain as phenomenological successful a FN model that features all the complex coefficients satisfying the relation $ 1- \Delta x < |c^u_{ij} |, |c^d_{ij}| < 1 + \Delta x \ \forall \, i,j=1,\dots,3$ , i.e. implementing in a specific way the idea that all the coefficients have to be similar in size. If spanning the entire $U(3)^3$ parameter space the criterion would not be met, then the model had to be discarded. From this point of view, the minimization of Eq.~\eqref{eq:FN_cost_function} ensures to perform this task efficiently.
In particular, if all the sizes of the 18 complex entries in $c^{u,d}_{ij}$ turn out to fall in the acceptance range chosen, then the minimization performed in Eq.~\eqref{eq:FN_cost_function} allows us to find an optimal point in flavour space, more precisely the set of values $\bar{\theta}_{1,2,3}^{\,Q,u,d}, \bar{\delta}_{1,2,3}^{\,Q,u,d}$, together with the inferred $\bar{\epsilon}$, for which the FN model considered is manifestly natural in reproducing the pattern of quark masses and mixing observed.
Of course, the specific range at the basis of the $\mathcal{O}(1)$ criterion for the solution of the flavour puzzle remains a subjective matter.

\subsection{The flavour tuning}
\label{sec:how_to_fit_data_flavour_tuning}
As anticipated in the Introduction, the criterion dictating a viable FN model may not be sufficient to claim for a satisfactory solution of the SM flavour puzzle. Indeed, even when the $\mathcal{O}(1)$ criterion may be met, it is still possible that parameters of similar size conspire in order to reproduce the measured value for the predicted observable.
% Example 2x2, 1 eigenvalues can be massless if det=0. No symmetry reason just tuning.
In such a case, there would not be a symmetry reason behind the solution of the flavour puzzle, but rather an unfortunate case of tuning of the parameters involved. 
This fact is in good analogy with the study case of the hierarchy problem in the EW sector: there, the well-known Barbieri-Giudice measure~\cite{Ellis:1986yg,Barbieri:1987fn} acts as a discriminant to establish the goodness of a natural UV completion of the SM, looking at logarithmic derivatives of some key observables, e.g., the $Z$ boson mass, with respect to the parameters of the BSM theory.

In the following, we wish to introduce a similar notion for what concerns the flavour problem, and the FN mechanism in particular.
%The parameter-space values found with this procedure may then be interpreted as the optimal point around where to investigate the criterion discussed in $\mathbf{3.}$. 
For the purpose, we can promote the 10 observables $\{O_{K}\}_{K=1,\dots,10}$ involved in Eq.~\eqref{eq:chi2_standard} to be functions of the 18 complex coefficients appearing in Eq.~\eqref{eq:FN_coeffs}, and then proceed defining the dimensionless quantity:
\beq
\label{eq:delta_fine_tuning}
\Delta_{\rm FN} \equiv \max_{K,i,j} \, |\delta_{K,ij}|  \, \quad , \quad \delta_{K,ij} \equiv \frac{c^{u,d}_{ij}}{O_K} \frac{\delta O_K}{\delta c^{u,d}_{ij}}  \quad , 
% \Big|_{(\epsilon, \theta_{1,2,3}^{\,Q,u,d}, \delta_{1,2,3}^{\,Q,u,d} \,)}
\eeq
where the notation above gives understood that $\delta_{K,ij}$ has to be computed for both the real and imaginary part of the 18 complex coefficients.
The underlying meaning of Eq.~\eqref{eq:delta_fine_tuning} should be clear at this point: for a given solution to the flavour puzzle where all $|c^{u,d}_{ij}|\sim \mathcal{O}(1)$, if a small variation in the real or imaginary part of the entries of $c^{u,d}_{ij}$ produces a change  in one of the observables at hand, one would end up with $\Delta_{\rm FN} > 0$; then, if the latter is greater than a certain threshold value quantifying fine tuning in flavour space, e.g. $\Delta_{\rm FN} > \mathcal{O}(10)$, the solution to the flavour problem found may be considered unnatural.

In our new approach, given an optimal point in flavour space represented by the set $\bar{\theta}_{1,2,3}^{\,Q,u,d}, \bar{\delta}_{1,2,3}^{\,Q,u,d}$ and $\bar{\epsilon}$, the computation of $\Delta_{\rm FN}$ in Eq.~\eqref{eq:delta_fine_tuning} would require already the evaluation of $2 \times 18 \times 10 = 360$ numerical derivatives, corresponding to the number of different $\delta_{K,ij}$.
Note that evaluating $\Delta_{\rm FN}$ at the optimal point found via the minimization of Eq.~\eqref{eq:FN_cost_function} is not sufficient to claim that the FN model considered is really unnatural. Indeed, the ideal approach would be to construct a fine grid in the 19 dimensional space and evaluate $\Delta_{\rm FN}$ and the size of the coefficients in each point of the parameter space. Such a procedure would allow to have a global assessment on the $\mathcal{O}(1)$ size of $c^{u,d}_{ij}$ and on the degree of fine tuning involved at the same time. Computationally, such a task would be very demanding for a single FN model, and essentially prohibitive for a comprehensive survey of FN models. 
 
In order to overcome this technical difficulty, we propose here an alternative way to encode the notion of fine tuning corroborated in Eq.~\eqref{eq:delta_fine_tuning}. The basic idea is to add a statistical weight to Eq.~\eqref{eq:FN_cost_function}, in order to optimize a new cost function that takes into account also the degree of fine tuning in flavour space. While such a strategy may be realized in several ways, in this work we considered the following modification to Eq.~\eqref{eq:FN_cost_function}:
\begin{equation}
\label{eq:FN_cost_function_fine_tuning}
\chi^2_{\rm FN, \alpha} = \chi^2_{\rm FN} + \alpha \sum_{K,i,j} \delta_{K,ij}^2 \ ,
\end{equation}
with coefficient $\alpha \geq 0 $ and, once again, with the contribution of real and imaginary part of $c^{u,d}_{ij}$ in the sum of the second term understood. 

Equipped with Eq.~\eqref{eq:FN_cost_function_fine_tuning} and the fine-tuning estimator reported in Eq.~\eqref{eq:delta_fine_tuning}, in our novel approach to the SM flavour puzzle one could imagine to proceed as follows.
First, one would perform the minimization of $\chi^2_{\rm FN, 0}$ in order to assess if the FN model under consideration were able to address the flavour puzzle according to the $\mathcal{O}(1)$ criterion attached to it. Then, one may compute $\Delta_{\rm FN}$ and compare it to the desired threshold of acceptance for fine tuning in flavour space: if $\Delta_{\rm FN}$ would result greater than the established threshold, the model should be considered fine-tuned. At that point, one would not reject immediately the model, but rather turn on $\alpha > 0$ and repeat the exact same procedure. One should minimize again Eq.~\eqref{eq:FN_cost_function_fine_tuning} taking this time into account also the contribution from the term involving $\delta_{K,ij}$, check on the size of $c^{u,d}_{ij}$, and check on the value of $\Delta_{\rm FN}$. Depending on the outcome, one could perform this exercise iteratively for increasing $\alpha$ values, and consider the model natural if, for some values of the parameter $\alpha$, both the $\mathcal{O}(1)$ criterion is respected and, at the same time, the value for $\Delta_{\rm FN}$ is found to be smaller than the fine-tuning threshold chosen. At the practical level, one can consider a finite set of $\alpha$ values logarithmically spaced in an interval with extremes $\alpha = 0$, and $\alpha = \bar{\alpha}$ such that $\chi^2_{\rm FN, \bar{\alpha}} \gg \chi^2_{\rm FN, 0}$. This may provide a pragmatic handle on the assessment of the degree of fine tuning for the FN solutions found.

\section{Charting FN models}
\label{sec:results}
In this section we present our comprehensive investigation on the FN mechanism based on the novel approach to the flavour problem detailed in the previous section. We highlight as the most interesting outcome of our study the existence of viable FN models with small $U(1)_X$ charge, that turned out to not suffer of fine tuning in flavour space. 
% We also briefly comment on how natural the flavon sector of these models should be retained.

\subsection{Explicit methodology}

Let us now give a more specific realization of the strategy highlighted in the previous section.
We start identifying the set of FN models that we are interested in. 

In our analysis, the expansion parameter $\epsilon$ is treated as an unknown and will be inferred from data. Hence, within our approach, specifying the set of charges determines completely the FN model under study. Recalling from Eq.~\eqref{eq:zero_charges} that the top-quark Yukawa value and baryon number conservation fix 2 out of the 9 FN charges, we need to fix the value of 7 FN charges in order to define a specific model. Allowing each charge to assume any integer number in the range $[-n,n]$, this naively implies that there exist $(2n+1)^7$ different configurations that would need to be inspected. However, the final number of independent models is actually lower. Indeed, the invariance under $\mathcal{G}_{F}$ of the gauge-kinetic term in the SM Lagrangian implies that, once a set of charges is assigned, the physics is invariant under permutations of the charges within a family. Therefore, since such permutations would simply correspond to a reordering of the quarks within the family without any physical implication, it is enough to select a specific ordering of FN charges to inspect all the physically different FN models. We adopt in our analysis the ordering:
\bea
\label{eq:X_ordering}
X_{Q_{1,2,3}}=\{a,b,0\}; \; X_{u_{1,2,3}}=\{c,d,0\}; \; X_{d_{1,2,3}}=\{e,f,g\}, \text{ with } \begin{cases} a\geq b \\ c\geq d \\ e \geq f \geq g \end{cases}.
\eea
One can then conclude that the number of possible charges for both $X_{Q_i}$ and $X_{u_i}$ actually is $(2n+1)(2n+2)/2$ each, while the combinations for $X_{d_i}$ are $(2n+1)(2n+2)(2n+3)/6$, with total independent charge assignments equal to  $(2n+1)^3(2n+2)^2(2n+3)/24$. 

In the present analysis we will consider the exploration of FN models up to $n=3$, scrutinizing in this way more than %65856 
65k different models. 

One important comment before moving on is in order: since $n_{ij}^{u,d}$ depend only on the absolute value of the difference between a FN-charges pair, see Eqs.~\eqref{eq:FN_nu}~-~\eqref{eq:FN_nd}, this implies that starting from a specific FN model and inverting the sign of all the charges will produce the ``mirror case'', corresponding to a distinct model in the UV, described by the same low-energy EFT. Mirror models, after reordering the charges according to Eq.~\eqref{eq:X_ordering}, would be already included in the set of models we aim to explore. Hence, it is sufficient to analyse only one of the two models in each mirror pair, further reducing the number of independent models that practically one has to consider.

The following step is the definition of the phenomenological input values. The dataset at hand to perform our study consist of the six quark masses, the three CKM mixing angles and the CP-violating phase. For the quark masses, we consider their running via the SM renormalization group up to the high energies probed at colliders; specifically, we adopt the results at 1~TeV given in ref.~\cite{Xing:2011aa}:
\bea
\label{eq:mass_inputs}
m_u/\GeV  & \ = \ & 1.17\cdot 10^{-3}\,, \quad\quad m_c/\GeV \ = \ 0.543\,, \quad\quad m_t/\GeV \ = \ 148.1\,, \no\\
m_d /\GeV & \ = \ & 2.40\cdot 10^{-3}\,, \quad\quad m_s/\GeV \ = \ 0.049\,, \quad\quad m_b/\GeV \ = \ 2.41\,.
\eea
We describe the CKM matrix via the standard parameterization~\cite{Chau:1984fp} and take at face value the outcome of the unitarity triangle analysis performed in ref.~\cite{Bona:2007vi}:\footnote{In particular, we adopt the result of the NP fit of ref.~\cite{Bona:2007vi} and neglect small effects related to the running due to weak interactions, see for instance~\cite{Antusch:2013jca}.}
\beq
\label{eq:CKM_inputs}
\sin \theta_{12}  = 0.22497\,, \quad\quad \sin \theta_{13} = 0.00368\,, \quad\quad \sin \theta_{23} = 0.04229\,, \quad\quad \delta = 65.9^\circ \,.
\eeq

Note that the underlying assumption of a flavour-breaking scale around the TeV may be suggestive in light of direct searches at the LHC, while being theoretically sound in relation to the hierarchy problem of the EW scale. At the same time, if we were assuming a breaking scale for the horizontal Abelian symmetry an order of magnitude greater, this would have only a marginal impact on the results obtained in our numerical analysis. In fact, taking the SM theory to be valid up to $\mathcal{O}(10)$~TeV~\cite{Xing:2011aa,Antusch:2013jca,Huang:2020hdv}, the values of the quark masses runned at that scale would differ at most of 10$\%$ with respect to the corresponding ones at $\mathcal{O}(1)$~TeV. On the other hand, the validation of the conclusions drawn from our analysis may possibly need to be revised if the dynamics responsible of the flavour problem were to be originated really far away from the EW scale, e.g, at the typical scale of Grand Unification, $\Lambda_{\rm GUT} \sim \mathcal{O}(10^{13})$~TeV. In such a scenario, assuming the SM to be the correct theory up to $\Lambda_{\rm GUT}$, one may find out the impact of the renormalization group on the runned quark masses to be much more important: The value of the top-quark mass, in particular, may no longer underlie $y_{t} \simeq 1$, and the relation between FN charges illustrated in Eq.~\eqref{eq:FN_charge_top} may be much less motivated. Furthermore, the playground for a  model-independent analysis as the one pursued in the present work may also cease to hold due to the possible relevance of mass-threshold effects of some new dynamics between $\Lambda_{\rm GUT}$ and the EW scale.

%Using the values from Eqs.~\eqref{eq:mass_inputs}~-~\eqref{eq:CKM_inputs}, we can now minimize Eq.~\eqref{eq:FN_cost_function} for all the models under scrutiny. Given the complexity of the parameter space, for each definite set of FN charges we perform a two-step minimization: we first employ the Basin-Hopping method~\cite{1998cond.mat..3344W}, and subsequently use the MIGRAD algorithm~\cite{1975CoPhC..10..343J}, randomly initialized by tens of trials, to cross-check the outcome of the minimization procedure. 

Using the values from Eqs.~\eqref{eq:mass_inputs}~-~\eqref{eq:CKM_inputs}, we can now minimize Eq.~\eqref{eq:FN_cost_function} for all the models under scrutiny. Given the complexity of the parameter space, for each definite set of FN charges we perform a two-step minimization.
At first, we employ the Basin-Hopping method, namely a global optimization technique where local minimization of the likelihood is supported by an acceptance test analogous to the Metropolis criterion employed in ordinary Monte Carlo algorithms, see~\cite{1998cond.mat..3344W} for more details. By construction, the Basin-Hopping algorithm allows us to reasonably tackle the hard problem of dealing a priori with several different local minima.
As a second step, we  also use the MIGRAD algorithm~\cite{1975CoPhC..10..343J}, the fast gradient-descent minimization with variable metric implemented in the MINUIT package, see e.g.~\cite{iminuit}. For each model, we have randomly initialized MIGRAD by hundreds of trials, in order to cross-check the outcome of the minimization procedure obtained with Basin-Hopping, further optimizing on the minimum previously found. Within this elaborated procedure, while checking on the size of each individual $\mathcal{O}(1)$ coefficient, we should have numerically coped also with possible issues arising in the presence of degenerate minima.

In the following, we adopt as $\mathcal{O}(1)$ criterion for the size of $c^{u,d}_{ij}$ the range of acceptance [0.4, 1.6], namely we will consider a deviation from unity of at most $\Delta x = 0.6$ for $|c^{u,d}_{ij}|$ to dub a certain FN model under consideration \textit{phenomenologically successful}.

Finally, for the assessment of the degree of fine tuning of the solutions found, we further minimize, when necessary, Eq.~\eqref{eq:FN_cost_function_fine_tuning} for $\log_{10}\alpha \in \{-6,-5,-4,-3,-2,-1,0\}$. From our numerical analysis we explicitly observe that $\alpha = 10^{-6}$ typically gives back an outcome identical to what obtained minimizing originally Eq.~\eqref{eq:FN_cost_function}, while setting $\alpha = 1$ makes the presence of the first term in Eq.~\eqref{eq:FN_cost_function_fine_tuning}
 irrelevant in the minimization procedure.

\subsection{Selected results}

We now analyse and summarize the results we obtained applying the procedure described in the previous sections. As a first observation, we report that we found a large number of phenomenologically viable FN models: this may come as a surprise if one considers that we are restricting our investigation to rather low charge assignments for the FN models in the UV, while the $\mathcal{O}(1)$ criterion under consideration should be considered quite restrictive a priori, allowing for a 60\% deviation from coefficients equal to unity.

We singled out $\sim$~650 models capable to address the flavour puzzle in the quark sector of the SM by means of coefficients $c^{u,d}_{ij}$ that show a natural size. This number of solution doubles, once we take into account for each of them the corresponding mirror solution. Hence, as one of the major highlights of this study, out of the $\sim$ 65k cases examined, we observe that about 1.3k FN models with $U(1)_{X}$ charge in absolute value $\leq3$ can actually account for the SM flavour puzzle with a natural range for $|c^{u,d}_{ij}|$.

\begin{figure}[!t]
\centering
\includegraphics[width=0.87\textwidth]{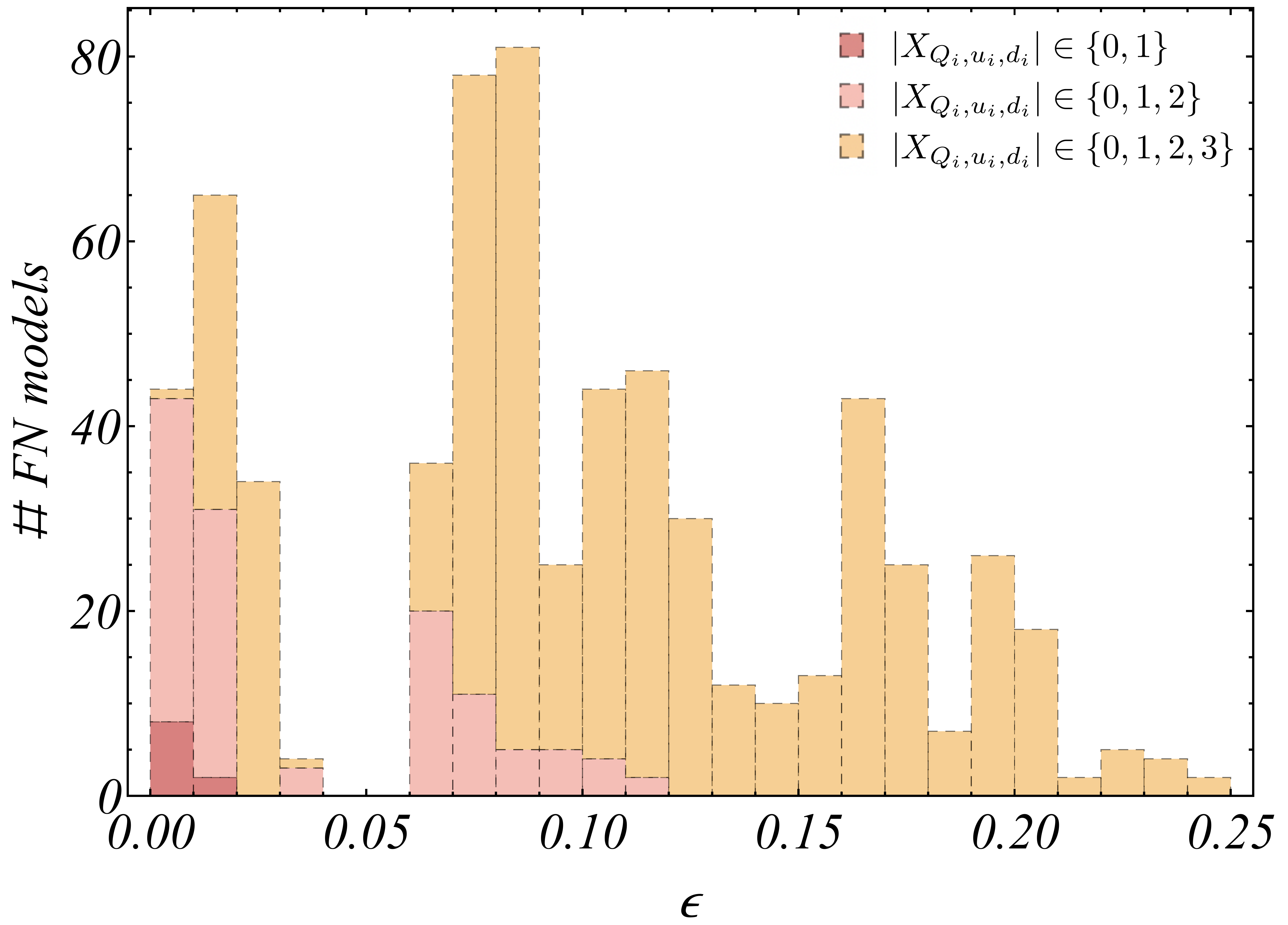} 
\caption{\it Histogram of the number of viable FN models with charges in absolute value $\leq 3$ as a function of the perturbative parameter $\epsilon = v_{\phi}/\Lambda$. The models are divided into three distinct classes according to their degree of ``minimality'': in red FN models have charges that are only 0 or $\pm 1$, in pink FN models have at least one charge that is equal to $\pm 2$, in orange the case where at least one charge gets equal to $\pm 3$. In our analysis, the $\epsilon$ parameter is inferred from data and ranges from a minimum value equal to 0.005 to a maximum close to 0.25. See text for more details.}
\label{fig:histoFN}
\end{figure}

In Fig.~\ref{fig:histoFN} we report the stacked histogram of the number of viable FN models as a function of the value for the perturbative parameter $\epsilon$, inferred from data. We divide the models in three separate classes, according to the values required for the charges of each model. In particular, we report in red the models where $X_{Q_i,u_i,d_i}$ only assume values of 0 or of $\pm 1$, in pink the models where at least one charge is equal to $\pm 2$, and in orange the remaining ones, i.e. where there is at least one FN charge equal to $\pm 3$. 

A few comments are then in order. First, we observe that the inferred values for $\epsilon$ are all in the range going from a minimum close to $0.005$ to a maximum around $0.25$. As one may have reasonably guessed, models described by small values of the charges $X_{Q_i,u_i,d_i}$ are correlated with a small FN perturbative parameter. Indeed, small FN charges will produce small entries in the matrices $n_{ij}^{u,d}$, hence requiring lower values for $\eps$ in order to reproduce the desired phenomenology. 
On the other hand, once we allow for larger values of the charges $X_{Q_i,u_i,d_i}$, i.e. larger entries for the matrices $n_{ij}^{u,d}$, a larger expansion parameter $\epsilon$ is consequently probed by data.
Note that the histogram of Fig.~\ref{fig:histoFN} hints for two prominent modes: one for $\eps \sim 0.01$ and another one for $\eps \sim 0.08$. Interestingly, for the small FN charges considered, the typical choice of $\epsilon \gtrsim 0.2$ often exploited in literature probes only the tail of the distribution captured in Fig.~\ref{fig:histoFN}.
It is also remarkable the fact that we find $\sim~$10 solutions featuring only $\{0,1\}$ for the FN charges in absolute value. These findings allow for the construction of the most economic FN theories possible in the UV, that to the best of our knowledge have never been considered in literature so far.

\begin{table}[!t]
\centering
\begin{tabular}{|ccc|ccc|ccc||c|}
\hline
\boldmath $ X_{Q_1} $ & \boldmath $ X_{Q_2} $ & \boldmath $ X_{Q_3} $ & \boldmath $ X_{u_1} $ & \boldmath $ X_{u_2} $ & \boldmath $ X_{u_3} $ & \boldmath $ X_{d_1} $ & \boldmath $ X_{d_2} $ & \boldmath $ X_{d_3} $ & \boldmath $ \epsilon $ \\[1mm]
\hline
0 & 0 & 0 &  1 & -1 & 0 & -1 & -1 & -1 & 0.005 \\
1 & 0 & 0 & -1 & -1 & 0 & -1 & -1 & -1 & 0.006 \\
1 & 0 & 0 &  0 & -1 & 0 & -1 & -1 & -1 & 0.006 \\
1 & 1 & 0 &  0 & -1 & 0 & -1 & -1 & -1 & 0.012 \\
%1 & 1 & 0 &  0 &  0 & 0 & -1 & -1 & -1 & 0.013 \\
\hline
0 & 0 & 0 &  1 & -2 & 0 & -1 & -1 & -2 & 0.006 \\
0 & 0 & 0 &  1 & -2 & 0 & -1 & -1 & -1 & 0.005 \\
0 & 0 & 0 &  1 & -2 & 0 &  2 &  1 & -1 & 0.006 \\
0 & 0 & 0 &  1 & -1 & 0 & -1 & -1 & -2 & 0.006 \\
0 & 0 & 0 &  1 & -1 & 0 &  2 &  1 & -1 & 0.006 \\
0 & 0 & 0 &  1 & 1 & 0 &  -1 & -1 & -2 & 0.005 \\ %
0 & 0 & 0 &  2 & -1 & 0 & -1 & -1 & -2 & 0.006 \\
0 & 0 & 0 &  2 & -1 & 0 & -1 & -1 & -1 & 0.005 \\
0 & 0 & 0 &  2 & -1 & 0 &  2 &  1 & -1 & 0.006 \\
1 & 0 & 0 & -1 & -2 & 0 & -1 & -1 & -2 & 0.008 \\
1 & 0 & 0 & -1 & -1 & 0 & -1 & -1 & -2 & 0.007 \\
\hline
\end{tabular}
\caption{\it A selection of viable FN models characterized by a small expansion parameter, $\epsilon \ll 0.1$. In the first 4 lines we report a subset of the models with $\vert X_{Q_i,u_i,d_i} \vert \in \{0,1\}$, reporting only combinations that underlie different $n_{ij}^{u,d}$ for such values of the charges; see text for more details. In the last 11 lines we present some of the viable FN  models singled out in our analysis with $\vert X_{Q_i,u_i,d_i} \vert \in \{0,1,2\}$.
\label{Tab:eps_1}}
\end{table}

Given the large number of viable models found in our analysis, in the following we focus on a few interesting cases that are illustrative examples of \textit{minimal FN constructions}. As a first class of selected models, we list in Table~\ref{Tab:eps_1} the 15 cases characterized by the lowest values found for the FN expansion parameter, i.e. $\epsilon \lesssim 0.01$. In particular, we report 4 models whose charges involve only $\{0,\pm 1\}$ assignments, and 11 more models for which charges can also be equal to $\pm 2$. It is worth to mention that, for the models where all $X_{Q_i}=0$, one obtains that $n^{u(d)}_{ij} \equiv |X_{u_j(d_j)} |$; therefore, for any given model in Table~\ref{Tab:eps_1} satisfying this requirement, not only the mirror case is viable and corresponds to a physically distinct UV model (obtained inverting the sign of all the set of FN charges reported), but one may further obtain physically different models in the UV, viable in the IR, just inverting the sign of any subset of the FN charges listed in the table, since they will be described by the same low-energy EFT. For instance, if one takes the model given in the first line of Table~\ref{Tab:eps_1}, one can easily construct the mirror model that, following the prescription of Eq.~\eqref{eq:X_ordering} for the ordering of the charges, reads:

\beq
\label{eq:example}
\renewcommand{\arraystretch}{1}
\setlength{\arraycolsep}{1pt}
\begin{pmatrix}
X_{Q_i}\\
X_{u_i}\\
X_{d_i}\\
\end{pmatrix}
=
\renewcommand{\arraystretch}{1}
\setlength{\arraycolsep}{1pt}
\begin{pmatrix}
\phantom{-}0 && \phantom{-}0 && \phantom{-}0\phantom{-} \\
\phantom{-}1 && -1 && \phantom{-}0\phantom{-} \\
\phantom{-}1 && \phantom{-}1 && \phantom{-}1\phantom{-} \\
\end{pmatrix} \ ;
\eeq
moreover, other six different physical models can be further obtained just changing the sign of one or more (but not all) of the five non-trivial FN charges assigned.

In a similar fashion, we report in Table~\ref{Tab:eps_2} a list of 15 cases characterized by values for the expansion parameter an order of magnitude larger, i.e. $\epsilon \sim 0.1$. For these cases, we selected 10 models whose charges are in the subset $\{0,\pm 1,\pm 2\}$, and 5 more models for which FN charges up to $\pm 3$ are considered. In Appendix~\ref{sec:example} we report an explicit example showing how, starting from one of the models listed in Table~\ref{Tab:eps_2} together with the explicit values for the dimensionless coefficient matrices $c^{u,d}$ and for the rotation matrices $V_{Q,u,d}$ obtained from the optimization of Eq.~\eqref{eq:FN_cost_function}, the desired values for quark masses and the CKM parameters are perfectly reproduced.

\begin{table}[!t]
\centering
\begin{tabular}{|ccc|ccc|ccc||c|}
\hline
\boldmath $ X_{Q_1} $ & \boldmath $ X_{Q_2} $ & \boldmath $ X_{Q_3} $ & \boldmath $ X_{u_1} $ & \boldmath $ X_{u_2} $ & \boldmath $ X_{u_3} $ & \boldmath $ X_{d_1} $ & \boldmath $ X_{d_2} $ & \boldmath $ X_{d_3} $ & \boldmath $ \epsilon $ \\[1mm]
\hline
1 & 1 & 0 & -2 & -2 & 0 & -2 & -2 & -2 & 0.098 \\
1 & 1 & 0 & -1 & -1 & 0 & -2 & -2 & -2 & 0.081 \\ %
1 & 1 & 0 &  0 & -2 & 0 & -2 & -2 & -2 & 0.094 \\
1 & 1 & 0 &  0 & -1 & 0 & -2 & -2 & -2 & 0.093 \\
%1 & 1 & 0 &  0 &  0 & 0 & -2 & -2 & -2 & 0.101 \\
2 & 1 & 0 & -2 & -2 & 0 & -2 & -2 & -2 & 0.109 \\
2 & 1 & 0 & -1 & -2 & 0 & -2 & -2 & -2 & 0.094 \\
2 & 1 & 0 &  0 &  0 & 0 & -2 & -2 & -2 & 0.094 \\
2 & 2 & 0 & -1 & -1 & 0 & -2 & -2 & -2 & 0.112 \\ %
2 & 2 & 0 &  0 & -2 & 0 & -2 & -2 & -2 & 0.109 \\
2 & 2 & 0 &  0 & -1 & 0 & -2 & -2 & -2 & 0.109 \\
%2 & 2 & 0 &  0 &  0 & 0 & -2 & -2 & -2 & 0.105 \\
\hline
0 & 0 & 0 &  3 & -3 & 0 & -2 & -2 & -3 & 0.104 \\
1 & 0 & 0 & -2 & -3 & 0 & -2 & -3 & -3 & 0.098 \\
1 & 1 & 0 & -2 & -3 & 0 & -2 & -2 & -3 & 0.100 \\
2 & 0 & 0 & -2 & -3 & 0 & -2 & -3 & -3 & 0.104 \\
2 & 1 & 0 & -2 & -3 & 0 & -2 & -2 & -2 & 0.104 \\
\hline
\end{tabular}
\caption{\it A selection of viable FN models characterized by a larger expansion parameter, $\epsilon \sim 0.1$. Similarly to what already done for Table~\ref{Tab:eps_1}, here we report in the first 10 lines some of the viable FN models with $\vert X_{Q_i,u_i,d_i} \vert \in \{0,1,2\}$, while in the last 5 lines we list  some of the solutions found with the largest charge assignment considered, namely $\vert X_{Q_i,u_i,d_i} \vert \in \{0,1,2,3\}$.
\label{Tab:eps_2}}
\end{table}

We close this section tackling possible fine-tuning issues in the FN solutions found. In order to address this point, we computed the fine-tuning parameter $\Delta_{\rm FN}$ defined in Eq.~\eqref{eq:delta_fine_tuning} for each of the $\sim 650$ viable models identified. 
%For the bulk of the analysed models, the tuning parameter is found to be at most $\mathcal{O}(10)$ directly inspecting the optimal point found minimizing Eq.~\eqref{eq:FN_cost_function}.
For roughly half of the analysed models, the tuning parameter is found to be at most $\mathcal{O}(10)$ from a direct inspection of the optimal point in parameter space found minimizing Eq.~\eqref{eq:FN_cost_function}.
%For a minority of the viable FN models obtained, we applied the iterative minimization procedure related to Eq.~\eqref{eq:FN_cost_function_fine_tuning} and discussed in detail in section~\ref{sec:how_to_fit_data_flavour_tuning}.
For the remaining viable FN models obtained, we applied the iterative minimization procedure related to Eq.~\eqref{eq:FN_cost_function_fine_tuning} and discussed in detail in section~\ref{sec:how_to_fit_data_flavour_tuning}.
%In the end, we found only $\sim 50$ FN models with a fine-tuning parameter $\Delta_{\rm FN} \gtrsim \mathcal{O}(10)$, but well within a threshold of an order of magnitude larger, $\Delta_{\rm FN} <10^2$.\footnote{None of these potentially fine-tuned models corresponds to the ones selected in Tables~\ref{Tab:eps_1}~-~\ref{Tab:eps_2}.} On general grounds, we can affirm that more than 90\% of the viable FN solutions found by our method is not susceptible of severe cancellations in the $U(3)^3$ flavour space, that would have signalled a fine-tuning issue in the resolution of the SM flavour puzzle.
In the end, we found only $\sim 100$ FN models with a fine-tuning parameter that remained well above the threshold $\Delta_{\rm FN} = 10^2$.

Interestingly, two thirds of these fine-tuned models exhibit systematically $\Delta_{\rm FN} \sim \mathcal{O}(10^3)$.\footnote{None of these potentially fine-tuned models corresponds to the ones selected in Tables~\ref{Tab:eps_1}~-~\ref{Tab:eps_2}.} The most relevant observables for these technically pathological cases, i.e. the observables responsible of a large fine-tuning estimator, always correspond to the quark masses $m_{u,d}$. This fact may be connected to the observation that most of these special cases belong to the class of minimal FN models for which the charge assignment implies the existence of a 0 eigenvalue in the limit $c^{u,d}_{i j} \to 1$. In these fine-tuned FN models the role of the $\mathcal{O}$(1) coefficients turns out to be of particular importance:  $c^{u,d}_{i j} \neq 1$ allow to avoid peculiar cancellations related to the presence of 0 eigenvalues, and critically increase the rank of the resulting up-quark and/or down-quark Yukawa matrices.

In other words, in models where the same FN-charge assignment characterizes the first two generations, the mass hierarchy among the latter arises in fact as a result of a matrix with rank approximately smaller than the maximal one. Consequently, these models are naturally more exposed to fine-tuning problems. However, we wish to stress that a careful evaluation of $\Delta_{\rm FN}$, including both the mixing with the third generation and the interplay of real and imaginary parts of the complex coefficients $c_{ij}^{u,d}$, must be always performed before dubbing these FN models as fine-tuned. A zoom on a model involving same FN charges in the first two generations, but with $\Delta_{\rm FN} \sim \mathcal{O}(10)$, is explicitly given in the appendix.

On general grounds, we can affirm that about 80\% of the viable FN solutions found by our method is not susceptible of severe cancellations in the $U(3)^3$ flavour space, that would have signalled a fine-tuning issue in the resolution of the SM flavour puzzle.

%\subsection{Flavon naturalness}

\section{Conclusions}
\label{sec:conclusions}
Today, the SM flavour puzzle may offer one of the most relevant clues on the footprints of BSM physics at low energy. In this work, we took this opportunity to explore the origin of mass and mixing hierarchies in the quark sector of the SM Lagrangian in the context of the FN mechanism.
As the main novelty of our study, we proposed a new method to evaluate whether a FN model can be considered viable from the phenomenological point of view. Our approach is bottom-up and intimately connected to the symmetries of the problem in the IR, namely the invariance of the SM quark gauge-kinetic Lagrangian under $U(3)^3$. It provides a rather general strategy that can be, in principle, replicated also for any other interesting proposal aimed at addressing the SM flavour problem.

With the focus of the present numerical investigation on the FN theory, we reviewed here its low-energy connotation in the EFT language. We exploited such formulation to apply our new approach to the flavour puzzle and  scrutinize in this way a large set of FN models with small charge values under the horizontal symmetry $U(1)_X$. In particular, we have systematically explored all FN models with $U(1)_X$ charges in the range $\{0,1,2,3\}$ in absolute value. The class of FN models considered are also characterized by a single flavon VEV $v_{\phi}$ and a FN-messenger scale $\Lambda$ via the expansion parameter $\epsilon = v_{\phi}/\Lambda$, that in our analysis can be directly inferred from data. 

Out of the $\sim$~65k scenarios analyzed, we found that $\sim$~1.3k FN distinct models in the UV can naturally reproduce the observed quark masses and the CKM mixing pattern. The FN perturbative ratio $\epsilon$ is found to lie in the range bracketed by $\sim0.005$ from below and $\sim 0.25$ from above, with the popular choice  $\epsilon \simeq 0.22$, related to the Cabibbo angle, probing only the tail of the distribution obtained in Fig.~\ref{fig:histoFN}, for the small FN charges considered. Remarkably, we have also found 10 solutions where the FN constructions actually feature a very minimal charge assignment of $\{-1, 0, 1\}$, see Table~\ref{Tab:eps_1}, providing to the best of our knowledge the most economic window to flavour model building in the UV. 
% Within an illustrative toy model, we have also discussed the degree of naturalness of the flavon sector in these minimal FN theories, arguing that the NP scale $\Lambda$ of the FN messengers should not be highly decoupled from the EW one.

Finally, in our work we have also introduced an estimator for fine tuning in flavour space, similar in spirit to the well-known Barbieri-Giudice measure for a natural theory of the EW scale. We noted that the vast majority of the minimal FN constructions inspected, namely about 80\% of the total ones considered, do not involve peculiar cancellations in flavour space in the resolution of the SM flavour puzzle.   

In light of the interesting outcome of the present work, future promising directions are foreseeable. One may include in our exact same setting a detailed study of the leptonic sector as well, that would open up also the quest for the origin of neutrino masses. It would be certainly instructive to extend the present analysis to the case of two-Higgs-doublet models, and take into account in this manner a study case closer to UV completions where supersymmetry will be manifest. Eventually, one may generalize the present analysis to the case of non-Abelian symmetries and build up a strong connection with top-down approaches typically adopted in the phenomenology of Grand Unified Theories. Finally, we wish to reiterate that the novel approach to the SM flavour puzzle presented in this paper could be insightful even in a context very different from the one we were able to frame within the FN mechanism, leaving for a future study a direct application of our method to the case of theories of extra-dimensions and of partial compositeness.

\appendix

\section{A worked-out example}
\label{sec:example}
In this Appendix we explicitly discuss how the quark masses, the three CKM mixing angles and the CP-violating phase can be reproduced, following the approach defined in this work. For the sake of definitiveness, we explicitly pick up one of the models listed in Table~\ref{Tab:eps_2}, 

\beq
\label{eq:example}
\renewcommand{\arraystretch}{1}
\setlength{\arraycolsep}{1pt}
\begin{pmatrix}
X_{Q_i}\\
X_{u_i}\\
X_{d_i}\\
\end{pmatrix}
=
\renewcommand{\arraystretch}{1}
\setlength{\arraycolsep}{1pt}
\begin{pmatrix}
\phantom{-}1 && \phantom{-}1 && \phantom{-}0\phantom{-} \\
-2 && -2 && \phantom{-}0\phantom{-} \\
-2 && -2 && -2\phantom{-} \\
\end{pmatrix}
\ \ , \ \ \epsilon = 0.0894772\,.
\eeq
In order to perform this task, we also need the coefficient matrices $c^u$ and $c^d$ relative to this specific model, %defined at Eq.~\eqref{eq:FN_rotation},
together with the rotation matrices $V_Q$, $V_u$ and $V_d$ obtained minimizing Eq.~\eqref{eq:FN_cost_function} in order to generate such coefficients. The coefficient matrices relative to the example described in Eq.~\eqref{eq:example} read

\beq
\label{eq:ex_cu}
c^u=
\renewcommand{\arraystretch}{1.5}
\setlength{\arraycolsep}{2pt}
\begin{pmatrix}
0.296996 - 0.962979i && -0.171824 + 0.987427i && -0.218137 + 0.965404i \\ -0.288732 + 0.962114i && 0.178763 - 0.987484i && 0.225587 - 0.995365i \\
-0.173489 + 0.992865i && 0.270735 - 0.967206i && -0.185740 + 0.820763i \\
\end{pmatrix}
\,,
\eeq

\beq
\label{eq:ex_cd}
c^d=
\renewcommand{\arraystretch}{1.5}
\setlength{\arraycolsep}{2pt}
\begin{pmatrix}
-0.444605 - 0.892952i && -0.479893 - 0.882641i && 0.0586865 + 1.00937i \\ 0.554765 + 0.830214i && 0.438668 + 0.892558i && -0.129055 - 0.983295i \\
-0.786523 - 0.605194i && -0.483583 - 0.866336i && 0.417669 + 0.895754i \\
\end{pmatrix}
\,,
\eeq
while the relative rotation matrices are found to be equal to 

\beq
\label{eq:ex_Vq}
V_Q=
\renewcommand{\arraystretch}{1.5}
\setlength{\arraycolsep}{2pt}
\begin{pmatrix}
0.712910 && 0.701254 && 0.00107484 + 0.0000491122i \\ 
0.690800 + 0.0611042i && -0.702053 - 0.0621101i && -0.148837 - 0.0131664i \\
-0.0766250 + 0.0703512i && 0.0789678 - 0.0725969i && -0.728127 + 0.668957i \\
\end{pmatrix}
\,,
\eeq

\beq
\label{eq:ex_Vu}
V_u=
\renewcommand{\arraystretch}{1.5}
\setlength{\arraycolsep}{2pt}
\begin{pmatrix}
0.706160 && 0.708052 && (-5.87950 -676.174i)\cdot 10^{-6} \\ 
0.220636 - 0.672735i && -0.220035 + 0.670942i && -0.00404038 + 0.0123197i \\
-0.00498184 - 0.00773051i && 0.00413596 + 0.00817755i && -0.497287 - 0.867489i \\
\end{pmatrix}
\,,
\eeq

\beq
\label{eq:ex_Vd}
V_d=
\renewcommand{\arraystretch}{1.5}
\setlength{\arraycolsep}{2pt}
\begin{pmatrix}
0.648493 && 0.109379 && 0.687503 + 0.307949i \\ 
0.419039 - 0.263742i && -0.787325 + 0.184842i && -0.306722 + 0.0819806i \\
0.576321 - 0.0464451i && 0.548613 + 0.181694i && -0.531034 - 0.222960i \\
\end{pmatrix}
\,.
\eeq
Inverting now Eq.~\eqref{eq:FN_rotation}, we observe that we can combine Eqs.~\eqref{eq:example}~-~\eqref{eq:ex_Vd} in order to obtain the Yukawa matrices
\beq
\hat{y}^u_{ij}= \(V_Q \ c^u \eps^{n^u} \ V_u^{\dagger} \)_{ij} \ \ , \ \  
\( V_{\rm CKM} \, \hat{y}^d \)_{ij} = \(V_Q\ c^d \eps^{n^d} \ V_d^{\dagger} \)_{ij} \ ,
\eeq
where the matrices $n^{u,d}$ are defined according to Eqs.~\eqref{eq:FN_nu}~-~\eqref{eq:FN_nd}. Squaring those matrices and multiplying them by $v_H^2/2$, one obtains the squared mass matrices:
\beq
\label{eq:ex_m2}
\left(m^u (m^u)^\dagger\right)_{ij}= \frac{v_H^2}{2}\(\hat{y}^u \ {\hat{y}^{u\,\dagger}} \)_{ij} \ \ , \ \
\left(m^d (m^d)^\dagger\right)_{ij}= \frac{v_H^2}{2}\(  V_{\rm CKM} \, \hat{y}^d   \hat{y}^{d\,\dagger} \, V_{\rm CKM}^{\dagger}  \)_{ij} \, .
\eeq
Diagonalising the above matrices, one can identify the (square of the) quark masses with the eigenvalues, $m_{u,c,t}^2$ and $m_{d,s,b}^2$, and eventually one can properly combine the eigenvector matrices to construct the CKM matrix. It is worth recalling that the CKM CP-violating phase $\delta$ suffers from convention choice; therefore, only the values of the CKM mixing angles $\theta_{12,13,23}$ can be directly extracted in an unambiguous way, related to the absolute value of the CKM matrix elements found. In order to reconstruct the CP-violating CKM parameter in a phase-convention independent manner, one can compute the Jarlskog invariant, $\mathcal{J}$, that can be obtained from Eq.~\eqref{eq:ex_m2} evaluating the commutator of the quark mass matrices, more precisely~(see, e.g., ref.~\cite{Hocker:2006xb}):
\begin{eqnarray}
 & \ \Im \  & \det \left( \left[ m^u (m^u)^\dagger,  m^d (m^d)^\dagger \right] \right)  = \\ \nonumber 
& \  =  \ & 2 \mathcal{J} \, (m_{t}^2-m_{c}^2) (m_{t}^2-m_{u}^2) (m_{c}^2-m_{u}^2)
(m_{b}^2-m_{s}^2) (m_{b}^2-m_{d}^2) (m_{s}^2-m_{d}^2) \ .
\end{eqnarray}

Using the numerical values from Eqs.~\eqref{eq:example}~-~\eqref{eq:ex_Vd}, we find that we can indeed reproduce the values for the six quark masses, the three CKM mixing angles, as well as the Jarlskog invariant $\mathcal{J} = \cos \theta_{12} \cos \theta_{13}^2 \cos \theta_{23}
\sin \theta_{12} \sin \theta_{13} \sin \theta_{23} \sin \delta \simeq 3.1 \times 10^{-5}$, with a per-mill level precision.

\acknowledgments{We thank Marco Ciuchini, Enrico Franco, Guido Martinelli, Ayan Paul, Arvind Rajaraman, Michael Ratz and Tim Tait for discussion. We are particularly grateful to Marco Nardecchia and Luca Silvestrini for relevant comments during early-stage collaboration. M.F and M.V. would like to express gratitude to the Mainz Institute for Theoretical Physics (MITP) for its hospitality and support during early stages of this project. M.F. is supported by the MINECO grant FPA2016-76005-C2-1-P and by Maria de Maetzu program grant MDM-2014-0367 of ICCUB and 2017 SGR 929.
The work of M.V. is supported by the NSF Grant No.~PHY-1915005.}

\bibliographystyle{JHEP}
\bibliography{bibliography}

\end{document}